\documentstyle[psfig]{lamuphys}
\makeatletter
\let\chapter\hid@chapter
\makeatother

\begin{document}
\setcounter{page}{1}
\pagenumbering{arabic}

\title{The Local Bubble, Local Fluff, and Heliosphere}

\author{Priscilla C. Frisch}

\institute{University of Chicago, Dept. of Astronomy \& Astrophysics,
5640 S. Ellis Ave., Chicago, IL  60637}

\maketitle

%%%%%%%%%%%%%%%%%%%%%%%%%%%%%%%%%%%%%%%%%%%%%%%%%%%%%%%%%%%
%
% Here follow my command definitions
%
%\newcommand{\la}{\mathrel{\hbox{\rlap{\hbox{\lower4pt\hbox{$\sim$}}}\hbox{$<$}}}}
%\newcommand{\ga}{\mathrel{\hbox{\rlap{\hbox{\lower4pt\hbox{$\sim$}}}\hbox{$>$}}}}

\begin{abstract}
The properties of the Local Bubble, Local Fluff
complex of nearby interstellar clouds, and the heliosphere are
mutually constrained by data and theory.
Observations and models of the diffuse radiation field, 
interstellar ionization,
pick-up ion and anomalous cosmic-ray populations, and interstellar
dust link the physics of these regions.
The differences between the one-asymmetric-superbubble and two-superbubble
views of the Local Bubble are discussed.  
\end{abstract}

\section{Introduction}

The Local Bubble, the Local Fluff complex of nearby interstellar clouds,
and the heliosphere, are three astronomical phenomena with properties 
that can not be determined independently.
The Local Bubble (LB) radiation field influences the ionization of both
nearby interstellar gas and the heliosphere.  The ionizations of nearby interstellar gas
and pick-up ions in the heliosphere constrain this radiation field.
The kinematics and properties of interstellar gas within and without the
heliosphere can be compared to set the confinement pressure of the
heliosphere and bow shock, and therefore the interstellar pressure.  Interstellar
grains observed within the solar system constrain the properties
of dust in interstellar clouds.  The morphology, properties and kinematics of nearby
interstellar gas must be consistent with models of the origin of the
soft X-ray background (SXRB).  

The LB, Local Fluff and heliosphere make a
science ``pyramid'', and the interrelation of the physical properties of these
three regions is shown in Fig. 1.  (The caption to Fig. 1
also summaries the abbreviations used in this paper.)
In this context, the local interstellar matter (LISM) is the same as the ``Local Fluff'' complex of interstellar clouds (LFC) within about 35 pc of the Sun.  

This conference, organized so beautifully by 
Dieter and his colleagues in Garching, is a timely sequel to two conferences
in the last two decades on interstellar gas in the heliosphere, and
on the local interstellar medium.\footnote{These two conferences
were the meetings on ``Interstellar Gas in Interplanetary Space, VI. MPAE Lindau Workshop'', June 187--20, 1980, 
held in Lindau, Germany, and ``Local Interstellar
Medium, IAU Colloquium No. 81'', held in Madison, Wisconsin June 4--6, 1984.}
\begin{figure}[tb!]
\psfig{file=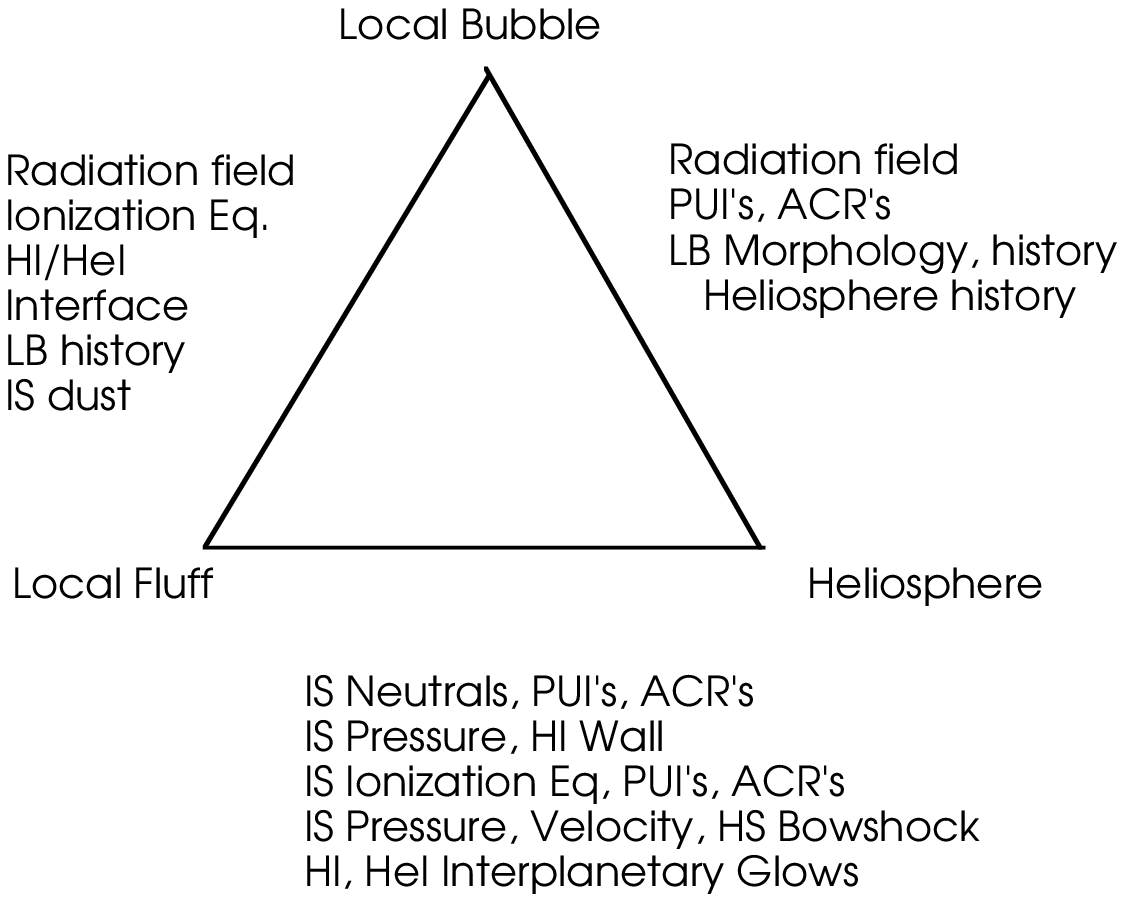,height=4.1in,clip=}
\caption[]{This figure shows the links between
the physics of the Local Bubble, Local Fluff, and Heliosphere.  
Abbreviations in the figure and text are:
CSSS=cloud surrounding the solar system; LFC=Local Fluff complex;
ISM=interstellar matter; LISM=local ISM; PUI=pick-up ions;
IS=interstellar; IE=ionization equilibrium;
ACR=anomalous cosmic rays; 
LB=Local Bubble; HS=heliosphere; 
SXRB=soft X-ray background; 
SCA=Scorpius-Centaurus Association. \vspace{-7 mm}}
\end{figure}
The initial linking of the Loop I superbubble, the LISM,
and the neutral interstellar gas detected within the solar system was presented in a series
of papers culminating in Frisch (1981).  
We used the Copernicus satellite to conduct spectral observations of the
interplanetary L$\alpha$ glow caused by the resonance fluorescence of
solar L$\alpha$ emission off neutral interstellar gas which had
penetrated the heliosphere\footnote{This is also known as the ``backscattered
radiation''.} (\cite{af77}).
I proposed these observations because I believed the cloud in front of $\alpha$ Oph
had invaded the solar system, causing the interplanetary L$\alpha$
glow.
The star $\alpha$ Oph is located at a distance of 14 pc in the direction of the intense radio continuum source, the North Polar Spur
\footnote{The North Polar Spur is the most intense section of the Loop I radiocontinuum superbubble feature.}.
My hunch evidently was right, and the velocities of the 
interplanetary glow emission feature projected to the direction of
$\alpha$ Oph (--22.3 km s$^{-1}$), and Na$^{o}$ and Ca$^{+}$ absorption
features seen in $\alpha$ Oph
(velocities --23.6 to --22.4 km s$^{-1}$ for the reddest
components, \cite{whk94}, \cite{cd95}) 
are nearly equal, although a possible velocity gradient is present in the
LFC.
At that time, the deceleration of interstellar gas at the heliopause was
unknown (c.f. \cite{lal93}).
The interstellar cloud fragment which surrounds the solar system is a member of the cloud complex seen in front of $\alpha$ Oph and other stars in the 
galactic center hemisphere of the sky.
The enhanced abundances of interstellar refractory elements
Ca$^{+}$, Fe$^{+}$ and Mn$^{+}$ seen in front of $\alpha$ Oph
led me to also conclude that the Sun resided in the edge of the Loop I
supernova remnant associated with the Scorpius-Centaurus stellar association.
The sightline towards $\alpha$ Oph exhibits the strongest interstellar
Ca$^{+}$ seen towards any nearby star.
The need to reconcile the asymmetric distribution of interstellar
gas within 30 pc of the Sun (e. g. \cite{fr96}, Table 1) and the symmetry of the ``bubble''
inferred to explain the SXRB led to my suggestion that the
Loop I supernova remnant had expanded asymmetrically into the low density
interarm region surrounding the solar system.  A different model forms the SXRB in an independent supernova
explosion around the Sun (unrelated to Loop I)
(e. g. \cite{dave}, Egger, this volume).
These contrasting views are reviewed in Breitschwerdt et al. (1996).
In this summary I will discuss:
\begin{itemize}
\item{the heliosphere as a probe of the cloud surrounding the solar system (CSSS) and the Local Bubble (LB)}
\item{the constraints LISM gas places on the LB by shadowing (or lack
thereof)}
\item{abundance and kinematical considerations, morphology and structure,
and the historical heliosphere.}
\item{My perspective of the LB.}
\end{itemize}
One point needs to be emphasized strongly.  There is an underlying question that must be answered, and that
question is ``What is the Local Bubble?''  The region of space commonly referred to as the ``Local Bubble'' coincides with the interior of Gould's
Belt and with the surrounding interarm
region, therefore the answer
to this question is not so obvious.  Fig. 2 compares the
distribution of interstellar clouds within 500 pc to the
space motions of the Sun and CSSS.  
I adopt the view of defining the LB by its walls (\cite{coxrey})).  Column densities N(H$^{o}$) of 10$^{19.8}$ cm$^{-2}$ and 10$^{20}$ cm$^{-2}$ attenuate the B and C band radiation, respectively, by a factor of
$\sim$3, giving a natural definition for the ``walls'' of the Local Bubble as the location where
N(H)$\geq$10$^{19.8}$ cm$^{-2}$, and the ISM opacity in the B-band exceeds unity.

\section{The Heliosphere as a Probe of the LISM and LB}

By way of background, I will summarize briefly the heliosphere (HS) structure and the properties of the CSSS.
The overall structure of the heliosphere appears to be a two-shock structure, with an inner
``termination shock'' where the solar wind goes from supersonic to
subsonic, the heliopause which is the stagnation surface between the
solar wind and interstellar plasma components, and a bow shock surrounding the
heliosphere.  The CSSS has properties T$\sim$7,000 K, n(H$^{o}$)$\sim$0.2 cm$^{-3}$ and n(p$^{+}$)=n(e$^{-}$)$\sim$0.1 cm$^{-3}$.  A magnetic field
is present, of unknown strength, but likely to be weak (B$\sim$1.5 $\mu$G, \cite{fr95}, \cite{gloeckler}).  
The relative velocity between the Sun and CSSS is 
$\sim$26 km s$^{-1}$, approaching from the direction l$\sim$5$^{o}$,
b$\sim$+16$^{o}$ in the rest frame of the Sun.  
In the rest frame of the Local Standard of Rest, this corresponds to a cloud moving towards us at 19--20 km s$^{-1}$ from
the direction l$\sim$335$^{o}$, b$\sim$--2$^{o}$.  \footnote{Note that 
since the Sun is immersed in this flowing cloud, and the Sun itself moves 
through space, the solar motion must be removed 
from the observed upwind direction in order to get the true space velocity of the CSSS in the Local Standard of Rest.}  Neutral interstellar atoms
cross the heliosphere into the solar system, and turn into the
pick-up ion population after ionization (by charge exchange with the
solar wind and photoionization) and capture by the solar wind.  The
pick-up ions are accelerated (perhaps at the termination shock of the
solar wind) and create the anomalous cosmic ray population, which 
propagates throughout the HS.

Heliosphere observations give
direct data on the physical properties of the interstellar cloud which surrounds the solar system,
the LFC, and radiation field within the LB.   
Listing the ways in which observations within the heliosphere
serve as useful probes of the LISM and LB: 

\begin{itemize}
\item{The CSSS feeds interstellar neutrals into the HS.  Thus, observations of pick-up ions and anomalous cosmic rays within the
heliosphere provide direct information on the ratios of interstellar
neutrals in the CSSS.  The pick-up ion (PUI) and anomalous cosmic ray (ACR) data can thus be used to constrain
interstellar ionization.  Because the ionization levels are sensitive to the radiation field, in turn the
PUI and ACR data prove to be a probe of the radiation field within the LB
(see Slavin and Frisch, this volume, and \cite{fs96}).
The elements He, Ne, H, O, N, C, and Ar have been observed in either the pick-up ion or anomalous cosmic ray populations.
Since C$^{o}$ is a subordinate  ionization state of carbon in the LISM, 
the C/O ratio in the PUI and ACR populations yield an estimate of the ionization of the CSSS (\cite{fr94}).}
\item{Observations of L$\alpha$ and 584 A backscattered radiation from interstellar H$^{o}$ and He$^{o}$ inside of the solar system,
respectively, yield information on the temperature,
velocity and density of the interstellar cloud which feeds neutrals into
the solar system (e.g. \cite{que}, \cite{flval97}, \cite{af77}, \cite{sfc97}).}
\item{Observations of L$\alpha$ absorption from the pile-up of interstellar H$^{o}$ outside of
the heliopause stagnation surface, due to the charge-exchange coupling
of interstellar H$^{o}$ and protons, constrain the fractional ionization
of the surrounding cloud and the Mach number of the bow shock of the solar
system.  The absorption from this pile-up must be included in L$\alpha$ 
profile fitting for absorption lines in nearby stars, for good D/H ratios.
In the $\alpha$ Cen direction,
an outer heliosphere model with Mach number=0.9, 
n(H$^{o}$)=0.14 cm$^{-3}$, n(e$^{-}$)=0.1 cm$^{-3}$, T=7600 K,
V=26 km s$^{-1}$ yields the best fit, to within the limited parameter
range considered, to observations of L$\alpha$ absorption towards
$\alpha$ Cen (\cite{gay}).}
\item{Interstellar dust observed by Ulysses and Galileo constrain the
mass distribution of interstellar dust grains in the CSSS, with a 
mean mass of 3 x 10$^{-13}$ gr corresponding to a grain radius of 0.3 cm$^{-3}$
for silicate density 2.5 gr cm$^{-3}$ (\cite{bag}).
}
\end{itemize}

\section{LISM Constraints on the Local Bubble and Heliosphere}

Models of the conductive interface between the LFC and LB plasmas, when
compared to PUI, ACR, and interstellar absorption line data, give direct information on the LB radiation field, LFC conductivity, magnetic field, and other physical 
quantities (\cite{sl89}, Slavin and Frisch, this volume).
The morphology, abundance patterns, ionization, density and temperature
of the cloud fragments which constitute the Local Fluff cloud complex give direct
information on the physical history of the LISM.  For instance,
the enhanced abundance patterns seen in the LFC,
in comparison to abundances seen in cold interstellar clouds, are interpreted
as indicating that nearby interstellar material has been processed
through a shock front (\cite{fr96}).  The velocity of the LISM gas
indicates an outflow of gas from the Scorpius-Centaurus Association (SCA),
suggesting, together with the anomalous abundances, that the LFC is
part of the superbubble associated with star formation in this region.
The flow of interstellar gas is seen in thirty-six interstellar Ca$^{+}$ absorption components seen in 17 nearby
stars yield a relatively coherent flow velocity of --0.1 $\pm$ 2.2 km s$^{-1}$
in a rest frame defined by the heliocentric velocity vector l=6$^{o}$.2,
b=+11$^{o}$.7, V=--26.8 km s$^{-1}$ (\cite{fr97}).  

The asymmetry of the LISM gas, within about 35 pc, belies the symmetry inferred for the local component of the SXRB.
Table 1 illustrates this well known property.
This asymmetry, and the fact that clouds in the LFC flow outwards from the SCA,
must be explained by any theory on the origin of the Local Bubble.
Note that because of this asymmetry, most of the mass within 35 pc
is in the galactic center hemisphere of the sky.

One important question:
Can the LFC shadow the 0.25 keV SXRB and
thereby give us information on the
spatial distribution of the emitting plasma?
Typical column densities to the ``edge'' of the LFC in several directions
are given in Table 1, where the edge distance is defined as N(H$^{o}$)/0.20.
The column densities are estimates in some cases, 
based on N(Ca$^{+}$)/N(H$^{o}$)=10$^{8}$, and an average cloud space
density of 0.20 cm$^{-3}$ is used.
From this, it can be seen that most of the interstellar gas in the LFC cloud complex has column
densities too low to provide significant shadowing of the SXRB -- i. e. well
below the 10$^{19.8}$ cm$^{-2}$ value needed to shadow the B-band.
\begin{table}[h]
\caption{Column Densities through Local Fluff}
\begin{center}
\begin{tabular}{ccccccc}
\hline \\
l,b(deg)&log N(H)$^{a}$&d(pc)$^{b}$&$~~~$&l,b(deg)&log N(H)$^{a}$&d(pc)$^{b}$\\
\hline \\
28,+15&18.86&10&&240,--11&17.9&1.1\\\
101,65&17.85&1.0&&289,--54&18.90&1.1\\
163,+5&18.24&2.7&&295,+46&18.53&4.8\\
214,+13&17.88&1.1&&330,--7&19.00&14\\
214,--60&18.70&0.7&&350,--53&18.49&4.4\\
\end{tabular}
\end{center}
$^{a}$ Log N(H)=log N(H$^{o}$+H$^{+}$).  See \cite{fw97}, \cite{fr97b}, \cite{nap} for original references.  $^{b}$d(pc) is the distance to the LFC edge,
which is calculated for n(H$^{o}$)=0.20 cm$^{-3}$.
\end{table}
\begin{figure}[tb!]
\psfig{file=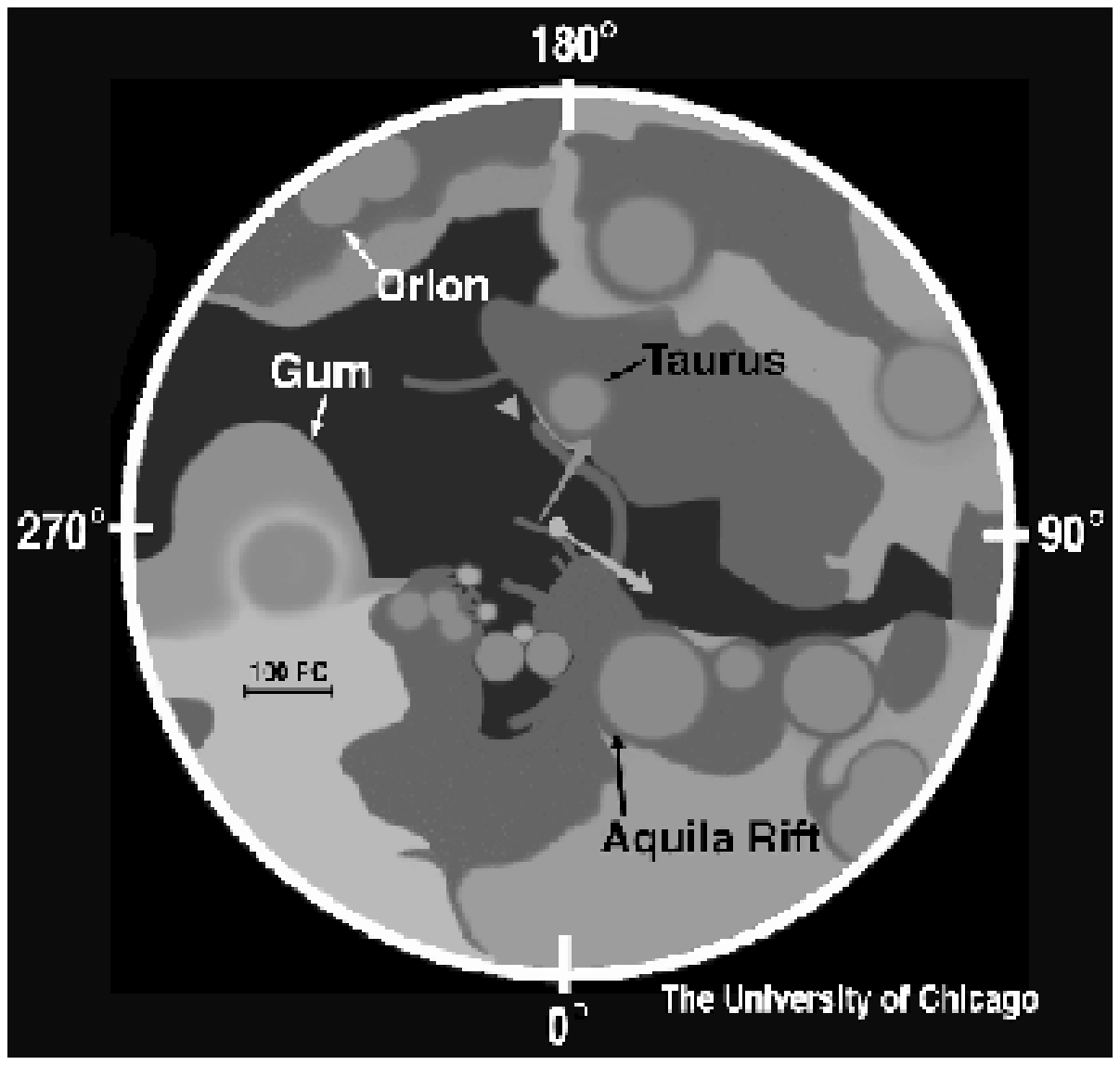,height=3.0in}
\caption[]{
The distribution of interstellar molecular clouds (traced
by the CO 1-$>$0 115 GHz rotational transition,\cite{dame}) and diffuse gas (traced
by E(B-V) color excess due to the reddening of starlight by interstellar dust \cite{lucke})
within 500 pc of the Sun are shown.  The round circles are molecular
clouds, and the shaded material is diffuse gas.  Interstellar matter is
shown projected onto the galactic plane, and the plot is labeled
with galactic longitudes.  The distribution of nearby interstellar matter
is associated with the local galactic feature known as ``Gould's Belt'',
which is tilted by about 15--20$^{o}$ with respect to the galactic plane.
Thus, the ISM towards Orion is over 15$^{o}$ below the plane, while
the Scorpius-Centaurus material (longitudes 300$^{o}$--0$^{o}$) is about
15--20$^{o}$ above the plane.  Also illustrated are the space motions of the
Sun and local interstellar gas, which are nearly perpendicular in the
LSR velocity frame.
The three asterisks are three subgroups of the
Scorpius-Centaurus Association.  The three-sided star is the Geminga
Pulsar.  The arc towards Orion represents the Orion's Cloak supernova
remnant shell.  The other arcs are illustrative of superbubble shells
from star formation in the Scorpius-Centaurus Association subgroups.  The
smallest (i. e. greatest curvature) shell feature represents the Loop I
supernova remnant.}
\vspace{-5 mm}
\end{figure}
The exceptions are the $\alpha$ Oph (l=36$^{o}$, b=+23$^{o}$, d=14 pc) and
HD149499B (l=330$^{o}$, b=--7$^{o}$, d=37 pc) sightlines, with
column densities logN(H$^{o}$)$\sim$19.57, $\sim$19.00 cm$^{-2}$, respectively. 
The $\alpha$ Oph column density is not well known.
In these directions 10\% -- 50\% percent attenuation of the B-band
emission may be expected.  These stars are in the direction of the 
dominant shadow (due to H$^{o}$ filaments composing the Loop I shell, \cite{cleary}) in front of the X-ray emission from Loop I. The star
HD149499B is located within $\sim$10$^{o}$ of the LFC flow direction in the
Local Standard of Rest.
Hardening of the SXRB is expected, in agreement
with the observations of a
dipole gradient pointed towards l=168$^{o}$.7, b=11$^{o}$.2
(\cite{snow}).  The plasma
emitting in the 0.10--0.18 keV region in the upwind hemisphere must be 
in front of the nearest 6 $\times$ 10$^{19}$ cm$^{-2}$ hydrogen column density.
This plasma could, therefore, be behind the upwind cloud with no significant
shadowing.  Alternatively, the prevalence of very small structure in the ISM
allows the possibility that the plasma and cooler clouds forming the
LFC are interspersed.

\section{Journey of the Sun through Space}

Our improved understanding of the morphology and kinematics of nearby ISM
in comparison to the space trajectory of the Sun permit a deeper understanding
of the historical changes in the galactic environment of the Sun, and the
effect those changes have on the heliosphere.
From Fig. 2, we see that within the past $\sim$100,000--200,000 years the Sun
emerged from the void of the surrounding interarm region and entered the
LF complex of clouds.  Within the past 10,000 years, and perhaps within the
past 2,000 years, the Sun appears to have entered the cloud in 
which it is currently situated (\cite{fr97}).  The physical properties
of these clouds constrain the configuration and properties of the
heliosphere.  It is notable that the space velocities of the Sun
and CSSS are nearly perpendicular, so that these ages are highly sensitive to
uncertain assumptions about cloud morphology and kinematics.

\section{Origin of the Local Bubble--One vrs. Two Bubbles}

What is the Local Bubble?  There is no agreement on the answer to this 
question.  Ask an X-ray astronomer and they will
probably tell you that it is the physical location of the 10$^{6}$ K plasma from
a recent local supernova explosion that emits radiation in the 0.1--0.18 KeV B-Band and 0.25 keV C-Band.  
Ask a radio astronomer, and they may be confused because the Sun
is located in an interarm region between the Orion spiral arm and
the Local Arm, which is a short ($\sim$1 kpc long) spur 
projecting from the Orion Arm.  Spiral arms are 
traced out by strings of molecular clouds and star-forming regions,
and Fig. 2 illustrates the nearby molecular clouds as defined by
observation of CO.  Interarm regions are regions with very low densities 
of interstellar matter.  

Historically, the LB concept is a mixture of
the view that a separate local supernova explosion formed the SXRB
(e. g. \cite{dave}, \cite{coxrey}) and the view that the superbubble formed by the
successive epochs of star formation in the Scorpius-Centaurus Association
have expanded asymmetrically into the low density interarm region
surrounding the Sun (\cite{fr81}, \cite{fr95}).  
The Davelaar et al. view has been updated by Egger and Aschenbach (1995, EA, also Egger, this volume),
who attribute the conventional H$^{o}$ filaments which bound the Loop I
supershell, and which are threaded by a displaced galactic magnetic field,
to a collision between the supernova around the solar system and the Loop I superbubble.

The salient properties of the asymmetric superbubble model for 
the LB are (see Frisch, 1995, for more details):
\begin{itemize}
\item{The three epochs of star formation (4--15 Myrs ago) in the SCA
each created superbubble structures, with the later structures evolving
within the cavities formed by the earlier events. In the asymmetric superbubble 
view these shells will have expanded asymmetrically into the low density interarm region surrounding the Sun.  The ISM surrounding the superbubbles was
initially asymmetric because of the local arm--interarm
configuration, and the Aquila Rift molecular cloud between l$\sim$20$^{o}$ and
l$\sim$40$^{o}$.}
\item{The Loop I supernova remnant, $\sim$250,000 years old (\cite{borken}), is confined because it expanded
into, and ablated material from, the Aquila Rift molecular cloud.  This can be
seen in the galactic interval l=17$^{o}$--27$^{o}$, b=0$^{o}$--10$^{o}$ when the configuration of the Aquila Rift CO cloud (\cite{dame}) is compared
with the narrow neck region of the North Polar Spur (\cite{sofue}).  The Aquila Rift molecular cloud is the
node region where all of the superbubble shells from the three epochs of 
star formation in the SCA, as well as the most recent supernova explosion creating
the North Polar Spur, converged after plowing
into the molecular gas and decelerating.  I propose here that the soft X-ray emission associated with the North Polar Spur itself occurs in a
position consistent with the formation of a galactic fountain with a footprint
in the disrupted Aquila Rift molecular cloud.}
\item{The characteristic filamentary
structure seen in the H$^{o}$ gas, which defines the annular ring attributed 
to the merged bubbles, appears to be due to confinement by the $\sim$5 $\mu$G 
magnetic field embedded in the filaments (\cite{cleary}) and does not require the explanation of
merging bubbles.
The superbubble shell boundaries portrayed in Fig. 2 represent 
the 21 cm filaments which are seen at negative galactic latitudes between l$\sim$40$^{o}$ and l$\sim$180$^{o}$, and which are threaded by the ambient magnetic field as is seen by Zeeman splitting and stellar polarization measurements.}
\item{The LFC is part of the expanding superbubble shell from the formation
of the Upper Scorpius subgroup 4--5 Myrs ago.  
Towards larger galactic latitudes, l=350$^{o}$ to l=40$^{o}$,
corresponding to the eastern boundary of Loop I,
the expansion of the shells was impeded by collision with the Aquila Rift
molecular gas.
At lower longitudes, or the western boundary of the Loop I superbubble,
expansion proceeded more freely and earlier shells expanded past the solar location,
reheating the low density gas in the anti-center interarm region.}  
\item{The Loop I supernova remnant has expanded inside
of the Upper Scorpius subgroup superbubble, and encountered denser ambient
gas than did the previous superbubble shells because of the proximity
to the denser Aquila Cloud.
The upwind direction in the Local Standard of Rest, 
l$\sim$335$^{o}$, b$\sim$--2$^{o}$, represents a direction offset
from the center of the Loop I supernova remnant by about 20$^{o}$.  }
\end{itemize}

The main source of disagreement between the asymmetric-superbubble versus symmetric-superbubble views is the requirement that in the latter scenario,
a separate supernova explosion in the anti-center hemisphere explains
the SXRB.  In the former view, the source of the SXRB is the ambient plasma
inside of the superbubble shells formed by the
first star formation epochs in the SCA
reheated by unspecified shocks and energetic radiation that would
propagate freely through the very low density material.  In the two-bubble
scenario, a coherent dense wall of neutral hydrogen at 40--70 pc, with N(H$^{o}$)$\sim$10$^{20}$ cm$^{-2}$
is postulated to separate the two bubbles, but there is no observational 
evidence for such a wall within 50 pc covering the central regions of the Loop I
bubble.  Counter-examples are easily found. 
For example, comparing the stars $\beta$ Cen (l=312$^{o}$, b=1$^{o}$, d=161 pc, log N(H$^{o}$)$\leq$19.5 cm$^{-2}$, \cite{frus}), and
HD 149499 B (l=330$^{o}$, b=--7$^{o}$, d=37 pc, log N(H$^{o}$)$=$19.0 cm$^{-2}$) show that over 30\% of the nearby gas is associated with the LFC.
The absence of X-ray emission from the Loop I interior may be due
to the fact this is an evacuated cavity.
An adequate model for the SXRB emission mechanism is required to establish
its origin (Sanders, this volume) and help resolve these differences.

The two scenarios agree on the distance of the H$^{o}$ 21 cm filaments, which 
are established by reddening measurements.  The LFC is in 
the interior of the ring, but it has distance $<$40 pc.
I believe the LFC is gas which is
part of an expanding superbubble shell from the SCA.
%
%\section{Acknowledgments}

This research has been supported by NASA grant NAGW-5061.

\end{document}